\documentstyle[12pt]{article}

\begin{document}

\newcommand{\be}{\begin{equation}}
\newcommand{\ee}{\end{equation}}
\newcommand{\bea}{\begin{eqnarray}}
\newcommand{\eea}{\end{eqnarray}}

Version 29th August 1996
\vfill

\centerline{\bf Collective Kondo effect in the Anderson--Hubbard lattice}
\bigskip
\centerline{P. Fazekas and K. Itai}
\medskip
\centerline{Research Institute for Solid State Physics}
\centerline{P.O. Box 49, Budapest 114,  H--1525 Hungary}

\bigskip

{\bf Abstract.}
The periodic Anderson model is extended by switching on a Hubbard $U$ for the  
conduction electrons. We use the Gutzwiller variational method to study 
the nearly integral valent limit. The lattice Kondo energy contains the 
$U$-dependent chemical potential of the Hubbard subsystem in the exponent, 
and the correlation-induced band narrowing in the prefactor. Both effects 
tend to suppress the Kondo scale, which can be understood to result from the 
blocking of hybridization. At half-filling, we find a Brinkman--Rice-type 
transition from a Kondo insulator to a Mott insulator.

\bigskip
\bigskip
Presented at the SCES '96 (Z\"urich, 1996)

To appear in Physica B

\vfill

\newpage

Recently, there has been  a great deal of interest in many-band models 
which describe hybridization between different subsystems of strongly 
correlated electrons. Schork and Fulde  introduced the problem of the 
Kondo-compensation of a magnetic impurity by a correlated band \cite{SF}. 
The corresponding periodic model can be envisaged either as an array of 
localized spins coupled to a Hubbard band (the Kondo--Hubbard lattice 
\cite{shi}), or as a periodic system of Anderson impurities hybridized 
with a Hubbard band (the Anderson--Hubbard lattice \cite{IF}). A particular 
version of the latter model was suggested for the remarkable correlated 
semiconductor FeSi \cite{FD}. We note that a closely related class of 
two-band models is the subject of intense investigation in connection with 
the resurgent interest in metallic ferromagnetism \cite{Tas,PS}. 

In contrast to their single-impurity counterparts, the Anderson--Hubbard (AH) 
and Kondo--Hubbard (KH) lattice models are not equivalent \cite{IF,Li}. 
In fact, they tend to lead to different predictions as to the 
effect of the conduction band Hubbard $U$ on the Kondo energy. A strong 
enhancement of the Kondo scale is found for the KH model, both for the 
single impurity \cite{KF} and the lattice \cite{shi}. As for AH models, 
there is at least one contribution which indicates the reduction of the 
Kondo scale with increasing $U$ for the impurity \cite{SF,TS}, and there 
is an argument for a reduced scale in the lattice \cite{IF}. The latter 
finding may be understood as the consequence of the blocking of hybridization 
processes. 
  
We consider the periodic Anderson--Hubbard model which describes an array of 
strongly correlated $f$-sites hybridized with a moderately strongly 
interacting $d$-band:
\bea
{\cal H} & = &  \sum_{{\bf k},\sigma}\epsilon_d({\bf k})
d_{{\bf k}\sigma}^{\dagger}d_{{\bf k}\sigma}  
+ \epsilon_f\sum_{{\bf j},\sigma}{\hat n}_{{\bf j}\sigma}^f
+ U_f\sum_{\bf j}{\hat n}_{{\bf j}\uparrow}^f
{\hat n}_{{\bf j}\downarrow}^f 
\nonumber \\[2mm]
& & +U_d\sum_{\bf j}{\hat n}_{{\bf j}\uparrow}^d{\hat n}_{{\bf j}\downarrow}^d
-v\sum_{{\bf j},\sigma}(f_{{\bf j}\sigma}^{\dagger}
d_{{\bf j}\sigma} + d_{{\bf j}\sigma}^{\dagger}f_{{\bf j}\sigma})
\label{eq:FD}
\eea
where ${\hat n}_{{\bf j}\sigma}^f=f_{{\bf j}\sigma}^{\dagger}
f_{{\bf j}\sigma}$, etc., the {\bf k} are wave vectors, and the {\bf j} are 
site indices. The $d$-bandwidth is $W$. In what follows, we take the 
strongly asymmetric Anderson model with $U_f\to\infty$ and the $f$-level 
$\epsilon_f<0$ sufficiently deep-lying so that we are in the Kondo limit: 
$1-n_f\ll 1$ where the $f$-valence is defined as 
$n_f=\langle\sum_{\sigma}{\hat n}_{{\bf j}\sigma}^f\rangle$.  The total 
electron density (per site, for one spin) is 
$n=\langle\sum_{{\bf j},\sigma}{\hat n}_{{\bf j}\sigma}^f+
\sum_{{\bf j},\sigma}{\hat n}_{{\bf j}\sigma}^d\rangle/2L$, where $L$ is the 
number of lattice sites. We will assume $1/2\le n\le 1$, so that there are 
enough electrons to fill at least the $f$-levels, and the $d$-band filling is 
variable up to half-filling. 

We use the Gutzwiller variational method, generalizing a previous treatment 
of the periodic Anderson model \cite{FB}. We postulate a Gutzwiller-projected 
hybridized band ground state 
\be
|\Psi\rangle  =  {\hat P}_{\rm G}^d\cdot{\hat P}_{\rm G}^f\cdot
\prod_{\bf k} \prod_{\sigma} [u_{\bf k}
f_{{\bf k}\sigma}^{\dagger} + v_{\bf k}d_{{\bf k}\sigma}^{\dagger}]|0\rangle
\label{eq:psi1}
\ee
where the mixing amplitudes $u_{\bf k}/v_{\bf k}$ are treated as independent 
variational parameters. The Gutzwiller projector for the $d$-electrons is
\be
{\hat P}_{\rm G}^d = \prod_{\bf g}[1-(1-\eta){\hat n}_{{\bf g}\uparrow}^d
{\hat n}_{{\bf g}\downarrow}^d]
\ee
where the variational parameter $\eta$ is controlled by $U_d$. For the 
$f$-electrons the full Gutzwiller projection is taken ${\hat P}_{\rm G}^f = 
\prod_{\bf g}[1-{\hat n}_{{\bf g}\uparrow}^f{\hat n}_{{\bf g}\downarrow}^f]$.
Here we consider only non-magnetic solutions corresponding to a mass-enhanced 
metal at $n<1$, or a renormalized-hybridization-gap insulator at $n=1$.

Omitting the details of the derivation (which relies on the Gutzwiller 
approximation), we quote our essential results. 
The expression for the optimized total energy density is
\bea
{\cal E} & = & \epsilon_f -n_d^0(1-n_d^0)q_d + U_d\nu_d^0
\nonumber \\[2mm]
& & -Wn_d^0\; q_d^0\cdot \exp{\left\{-\frac{\mu_0(U_d)-
\epsilon_f}{4(v^2/W)}\right\} }.
\label{eq:ekon}
\eea
The first line gives the energies of decoupled $f$-, and $d$-electrons. The 
coupling between the two subsystems is described by the last term which we 
identify as the Kondo energy of the Anderson--Hubbard model. The same 
Kondo scale can be identified in the deviation of the valence from 1
\be
1-n_f = \frac{n_d^0\; q_d^0}{4(v/W)^2}\cdot \exp{\left\{-\frac{\mu_0(U_d)-
\epsilon_f}{4(v^2/W)}\right\} }.
\label{eq:1mnf}
\ee

In the above equations, $n_d^0=n-1/2$ is the $v=0$ value of the conduction 
electron density (per spin). $\nu_d^0$ is double occupation, $q_d^0$ is 
the $q$-factor (the renormalization 
factor of the hopping amplitude) taken with 
$n_d^0$, and $\mu_0(U_d)$ 
is the chemical potential of the Hubbard band calculated (in the 
Gutzwiller approximation) for band filling $n_d^0$.  

Eqs. (\ref{eq:1mnf}) and (\ref{eq:ekon}) are as yet formal results: they 
express the solution in terms of the optimized parameters of the Hubbard 
subsystem. There is no closed-from result away from half-filling; the results 
of small-$U_d$, and of large-$U_d$, expansions are given in \cite{IF}. Here 
we discuss first the general features, and then the special case of 
half-filling.
 
The conduction band $U_d$ appears in the characteristic Kondo scale in two 
ways: through the prefactor and the exponent. The prefactor describes the 
correlation-induced narrowing of the $d$-band. The exponential factor can 
be written as $\exp{(-W/J_{\rm eff})}$ where we introduced the effective Kondo 
coupling
\be
J_{\rm eff} = \frac{4v^2}{\mu_0(U_d)-\epsilon_f}\; .
\label{eq:effj}
\ee
It differs from the Anderson lattice result \cite{FB,RU} inasmuch as 
$\mu_0(0)=(n_d^0-1/2)W$ is replaced by the $U_d$-dependent chemical potential 
$\mu_0(U_d)$. Comparing to the single impurity result \cite{SF}, we call 
attention to the extra factor of 2 in the numerator: the 
`lattice enhancement of the Kondo effect' \cite{RU} works also in the 
present model.

Due to the combined effect of the prefactor and the exponent, the lattice 
Kondo energy is a decreasing function of $U_d$. To make this finding 
plausible, one can argue that increasing $U_d$ suppresses charge 
fluctuations, and therefore blocks the hybridization processes which give 
rise to the Kondo coupling, and thus to the Kondo effect. The blocking of 
hybridization is carried to an 
extreme in the case of exact half-filling, as we discuss below:

According to a usual Luttinger's theorem argument, the half-filled case 
$n=1$ belongs to a non-magnetic insulator. The well-known  
$q_d=1-(U_d/2W)^2$ holds right up to $U_d=2W$ where the $d$-band undergoes 
a Brinkman--Rice transition \cite{BR}. However, in contrast to the usual 
one-band Brinkman--Rice transition which is a metal--insulator transition, 
here we have to do with an insulator--insulator transition. For $U_d<2W$, the 
system is a new kind of Kondo insulator with a renormalized hybridization 
gap \cite{IF}. For $U_d>2W$, the system is integral valent ($n_f=1$), the 
Kondo effect is completely quenched, and we should speak about a Mott 
insulator. This is certainly what we expect within the Gutzwiller 
approximation. Though the well-known criticism concerning the complete 
suppression of polarity fluctuations can be raised here, we note that up 
to the Brinkman--Rice transition, the Gutzwiller result is in a surprisingly 
good agreement with the predicted high-dimensional behaviour of the Hubbard 
model \cite{GMK}. The variationally derived lattice enhancement of the Kondo 
scale is at least qualitatively confirmed by a similar calculation for 
the Anderson lattice \cite{Jar}. Therefore we expect that our results which 
combine Hubbard and Anderson lattice features, have a similar justification 
for $U_d<2W$. For $U_d>2W$, a treatment based on replacing the Hubbard band 
with a $t$--$J$  \cite{MCC} or a Heisenberg model \cite{ITKF} may be more 
appropriate.

Our trial state is the lattice generalization of a lowest-order 
Varma--Yafet state \cite{VY} which is a potential source of shortcomings. 
Work on the impurity case is in a more advanced state  and indicates that 
processes which enhance the Kondo scale, may predominate \cite{KF,TS}. It 
has been argued \cite{TS} that to capture this effect variationally, one 
should consider states which are higher in the Varma--Yafet hierarchy. If a 
similar statement proves true for the lattice, our work served to demonstrate 
the limitations of the Ansatz (\ref{eq:psi1}). However, we believe that the 
lattice case may be different, for the following reason: By postulating a 
large Fermi surface \cite{ShF}, states from the entire Brillouin zone are 
used, and therefore (\ref{eq:psi1}) does contain contributions from 
electron--hole pairs with respect to the small (conduction electron) Fermi 
surface.    

{\bf Acknowledgement}. P. F. gratefully acknowledges financial support by 
the Hungarian National Science Research Foundation grant OTKA T-014201.

\end{document}